\documentclass[journal]{IEEEtran}

\ifCLASSINFOpdf
\else
   \usepackage[dvips]{graphicx}
\fi
\usepackage{url}
\usepackage{amsmath}
\usepackage{graphicx}
\usepackage{placeins}
\usepackage{float}
\usepackage{bbm}
\usepackage{subfigure}
\usepackage{verbatim}
\usepackage{amssymb}
\usepackage{amsmath}
\usepackage{amsthm}
\usepackage{mwe}
\usepackage{docmute}
\usepackage[ruled, vlined]{algorithm2e}

\usepackage[disable]{todonotes}

\usepackage{graphicx}

\begin{document}

\title{Adversarial Multi-Player Bandits for Cognitive Radar Networks}
% Note: 6 page max
\author{William W. Howard, Anthony F. Martone, R. Michael Buehrer
\thanks{W.W. Howard and R.M. Buehrer are with Wireless@VT, Bradley Department of ECE, Virginia Tech, Blacksburg, VA, 24061. \\
e-mails:$\{$wwhoward, buehrer$\}$@vt.edu  }
\thanks{A.F. Martone is with the U.S. Army Research Laboratory, Adelphi, MD 20783. (e-mail:{anthony.f.martone.civ}@army.mil).}
\thanks{The support of the U.S. Army Research Office (ARO) is gratefully acknowledged. }}% \vspace{-.6cm}}
% \thanks{The support of Applied Signals Intelligence, inc. is gratefully acknowledged. }\vspace{-.6cm}}

\maketitle
\pagenumbering{roman}

%%%%%%%%%%%%%%%%%%%%%%%%%%%%%%%%%%%%%%%%%%%%%%%%%%%%%%%%%%%%%%%%%%%%%%%%%%%%%%%%%%%%%%%%%
\begin{abstract}
    We model a radar network as an adversarial bandit problem, where the environment pre-selects reward sequences for each of several actions available to the network. 
    This excludes environments which vary rewards in response to the learner's actions. 
    Adversarial environments include those with third party emitters which enter and exit the environment according to some criteria which does not depend on the radar network. 
    The network consists of several independent radar nodes, which attempt to attain the highest possible SINR in each of many time steps. 
    We show that in such an environment, simple sub-band selection algorithms are unable to consistently attain high SINR. 
    However, through the use of adversarial multi-player bandit algorithms, a radar network can continue to track targets without a loss in tracking precision. 
\end{abstract}
%%%%%%%%%%%%%%%%%%%%%%%%%%%%%%%%%%%%%%%%%%%%%%%%%%%%%%%%%%%%%%%%%%%%%%%%%%%%%%%%%%%%%%%%%

%%%%%%%%%%%%%%%%%%%%%%%%%%%%%%%%%%%%%%%%%%%%%%%%%%%%%%%%%%%%%%%%%%%%%%%%%%%%%%%%%%%%%%%%%
\section{Introduction}
Multi-armed bandit models are an easy and natural tool for the analysis of cognitive radar. 
Cognitive radars follow the perception-action cycle, in which they iterate between making a decision and sensing the environment\footnote{When we refer to the environment, we specifically mean any targets and emitters aside from the radar network}, with the goal of improving the quality of future decisions.  
Multi-armed bandits (MABs) \cite{MAL-068, bandits, EXP3} follow the same cycle, where after taking an action, a player receives a reward from the environment, then chooses another action. 
In both formulations we have a finite number of actions available, which are all known to the player. 
Further, the rewards observed by MABs can be tailored to the cognitive radar scenario.

MABs have been used variously in the literature to analyze different cognitive radar scenarios. 
The authors of \cite{thornton2020efficient} consider coexistance scenarios for cognitive radar and cellular communications, implementing several bandit algorithms. 
Then in \cite{9455255}, the authors consider both stochastic and adversarial \emph{contextual} bandits, which aim to introduce vectors of contextual information to improve the learner's decisions. 
An automotive radar problem is considered in \cite{ferdowsi2019cyber}, where the authors aim to implement multi-armed bandit algorithms to improve security in autonomous connected vehicles.

In this work, we consider the behavior of a cognitive radar \emph{network}, i.e. a group of independent cognitive radar nodes \cite{Martone_CRN_loop}, through the lens of \emph{multi-player} multi-armed bandit algorithms (MMABs). 
Cognitive radar nodes are independent devices capable of monitoring the spectrum and modifying radar waveforms.

MMABs have been studied variously in the literature. 
In \cite{pmlr-v48-rosenski16}, the authors consider an algorithm called Musical Chairs, wherein players decide actions in a decentralized manner. 
The authors of \cite{mehrabian20a} go on to design algorithms tailored to an environment where the mean rewards for each arm and each player can vary. 
Since we're relating rewards to SINR, and assuming that interferers are of high enough power to impact the entire network, this complexity is not needed. 
Another algorithm is developed in \cite{NIPS2018_7952} which follows similar assumptions. 
Further, in \cite{NIPS2019_9375}, the authors describe a technique through which players can use \emph{implicit} communications to exchange information. 
This does not violate an assumption of no dedicated communications, since the players only use their actions to develop coordination.

Cognitive radar networks (CRNs) extend the capability of a single cognitive radar by placing many \emph{radar nodes} (also referred to as nodes) throughout an area.
This comes at the cost of resource management: the CRN will need some strategy to mitigate \emph{mutual interference}\footnote{Mutual interference occurs when one node in a CRN causes harmful interference to another node in the same network. } within the network. 
The strategy can be as simple as implementing a ``central node'' to determine the actions taken by each node. 
% This, however, is somewhat naive since we'd end up with a communications problem, as well as immediate issues if the central node were to be disabled. 
This, however, is somewhat naive, as the central node needs some method of selecting actions for each node and communicating them to the network. 
In this work, we instead consider a CRN using \emph{distributed cognition}, a term which implies completely independent nodes. 
Further, we make the assumption that there is no dedicated communication between nodes. 
Removing the dependence on dedicated communications removes the possibility of an outage. 
% This means that there is no chance to use explicit communication to exchange information on the time scale of individual actions. 

Bandit models typically include one of environment classes. 
Stochastic environments are those that provide rewards drawn from arbitrary distributions with some fixed mean. 
Adversarial environments, on the other hand, are those that pre-select some sequence of rewards. 
In addition, adversarial environments can choose the reward sequence with a priori knowledge of the algorithm being used by the player(s). 
Importantly, these environments are unable to modify this reward sequence after the game has begun.

\paragraph{Contributions} We demonstrate the use of multi-player multi-armed bandits in \emph{adversarial} reward scenarios, which have not been explored before in the cognitive radar application. 
We develop a system model for this scenario and discuss the benefits of a CRN which does not have centralized control. 
We support our arguments with results that demonstrate the performance of the MMAB algorithms, as well as the theoretical detection and tracking performance of a CRN implementing these technologies. 

\paragraph{Notation} We use the following notation. 
Matrices and vectors are denoted as bold upper $\mathbf{X}$ or lower $\mathbf{x}$ case letters.
Functions are shown as plain letters $F$ or $f$. 
Sets $\mathcal{A}$ are shown as script letters. 
The logical negation of a statement $a$ is given by an overline $\overline{a}$. 
The transpose operation is $\mathbf{X}^T$. 
The backslash $\mathcal{A}\backslash \mathcal{B}$ represents the set difference. 
Functions which indicate whether a value $x$ is present in a set $\mathcal{A}$ are denoted as $\mathbbm{1}_\mathcal{A}(x)$. 
The set of all real numbers is $\mathbb{R}$ and the set of integers is $\mathbb{Z}$. 
The speed of electromagnetic radiation in a vacuum is given as $c$. 

\paragraph{Organization} The rest of this work is organized as follows. In Section II, we discuss the system model, algorithms, and assumptions we use. 
% In Section III we detail the algorithms being used, and what assumptions we need to place on our radar network. 
In Section III we provide supporting results, and in Section IV we discuss our conclusions. 
%%%%%%%%%%%%%%%%%%%%%%%%%%%%%%%%%%%%%%%%%%%%%%%%%%%%%%%%%%%%%%%%%%%%%%%%%%%%%%%%%%%%%%%%%

%%%%%%%%%%%%%%%%%%%%%%%%%%%%%%%%%%%%%%%%%%%%%%%%%%%%%%%%%%%%%%%%%%%%%%%%%%%%%%%%%%%%%%%%%
\section{Background}
Broadly, multi-armed bandit models consider a game between a player and an environment. 
In time step $t$, which is one of many, the player $r_i$ chooses $a \in \mathcal{A}$, one of $M$ available arms\footnote{
Arms are also referred to as actions in the literature. 
Here, we will restrict ourselves to calling them arms, or later, sub-bands. }. 
The environment responds by providing a reward  $y_i(t)$ to the player based on a reward  function. 
The player can also choose to \emph{rest} in a given round, which results in a reward of zero. 
For bandit algorithms, rewards are often drawn from some arbitrary distribution, although no assumption on this distribution is needed save for the existence of a mean. 
An example of Gaussian rewards is given by Eq. (\ref{eq:ex_reward}). 
\begin{equation}
    \label{eq:ex_reward}
    y_i(t) \sim \mathcal{N}(\mu_a(t), \sigma_j^2) \;\; \text{ s.t. } y_i(t)\in[0,1]
\end{equation}
Note that we truncate rewards to the unit interval.
The player can then use the observed reward  to inform future decisions.

We can also consider the interaction between several ($N<M$) identical players in the form of multi-player MAB algorithms. 
Here, in each time step, each player $r_i, 1 \leq i \leq N$ selects an arm $a \in \mathcal{A}$ and observes the corresponding reward. 
If any two players $r_i, r_j$ select the same arm $a$, we say a collision has occurred, and each player observes a reward of $0$. 
Equivalently, a collision can be thought of as two cognitive radar nodes occupying the same spectrum resource simultaneously, which causes high levels of interference at both receivers. 
This is where the ability of a player to rest becomes attractive: rather than colliding in a given round, any player can choose to rest. 
Collisions are observed by each player $i$ through an indicator function
\begin{equation}
\label{eq:collision_indicator}
    c_i(t) = \mathbbm{1}_{\mathcal{E}'}(i)
\end{equation}
where $\mathcal{E}'$ is the set of colliding nodes in each time step. 
We assume that the environment generates this function for each node.

Players which rest will still receive a zero reward, but if this results in another player choosing an arm and not colliding, the network's cumulative reward will still increase. 
Note that this effect is only on average, and each node will be unaware of the cumulative reward. 
However, by making this option available, the tracking performance of a network is increased. 
We can now represent the reward function as
\begin{equation}
    \label{eq:ex_reward_net}
    y_i(t) \sim \mathcal{N}(\mu_a(t), \sigma_j^2) * c_i(t) \;\; \text{ s.t. } y_i(t)\in[0,1]
\end{equation}
to denote the dependence on collisions.

%Further, we'll assume that the network we're designing has \emph{no central coordinator or fast-time\footnote{Fast-time is used to denote to represent any activities on the pulse-to-pulse scale} communication}. 
Further, we will assume no central controller. 
This means that each node is functionally independent and makes its own decisions according to some algorithm, rather than being explicitly controlled by another node or system. 
This implies that any strategy executed by the nodes must occur in a distributed manner. 
In addition, we make no provision for a dedicated communications channel; nodes are not connected (by wire or wireless means) to any other node. 
We will however allow for infrequent communication from each node to some fusion center where the distributed observations from the network can be combined.

Note that the network we are describing is homogeneous; each node is identical in function. 
This means that whatever algorithm we implement will be executed by each node. 
The above assumption of no central coordinator is, in effect, a statement of this homogeneity. 
This does not prohibit the network developing different functions at each node, and in fact, we will see an algorithm (Coordinate \& Play) that selects one node to suggest actions to the network. 
This is compatible with the above assumptions, since the network is still homogeneous and each node is capable of performing this function.

Adversarial environments are those that have a priori knowledge of the algorithm being implemented by the network, and the ability to pre-select a reward sequence for each action the network can take. 
They do not however have the ability to modify the reward sequence once the game is initiated. 
In the example of Gaussian rewards, this means that the environment can pre-select a sequence of rewards $\mu_j(t)$ for each arm and each time step in the game. 
The objective of the player(s) then is to maximize the cumulative observed reward, summed over both time and the players, as defined in Eq. (\ref{eq:cum_reward}). 
\begin{equation}
    \label{eq:cum_reward}
    X_T = \sum_{t=1}^T \sum_{i=1}^N y_i(t)
\end{equation}
In the context of this work, we model an environment with interference powers that change infrequently and on a pre-set schedule that is unknown to the network.

The performance of any given algorithm receiving cumulative reward $X_T$ in time step T can be compared against the cumulative reward observed by an omniscient algorithm, which in each time step chooses the combination of actions with the highest reward, and always observes the mean reward for the action chosen. 
Let $\mu^*(t) = \max_{a_j \in \{a\}} \mu_a(t)$ be the action \emph{in each time step} with the highest mean reward. 
Since the reward is not constant for each arm we need to be specific. 
We can then define the \emph{expected regret} Eq. (\ref{eq:ex_regret}) \cite{MAL-068}: 
\begin{equation}
    \label{eq:ex_regret}
    R_T = \sum_{t=1}^T \mu^*(t) - \mathbb{E}\left[ \sum_{t=1}^T \sum_{i=1}^N y_i(t) \right]
\end{equation}
which is the difference between the omniscient algorithm's cumulative reward and the \emph{expected} reward of the algorithm being evaluated. 
Note that neither the cumulative observed reward Eq. (\ref{eq:cum_reward}) nor the expected regret Eq. (\ref{eq:ex_regret}) are ever known to the radar nodes. 
In addition they are never aware of the optimal arm or the true arm means. 
% Note that the expected regret is not a quantity ever known to the learner; neither the optimal arm nor the true arm means are ever known. 

We can design a radar network to fit this bandit formulation nicely. 
Simply let each player correspond to a radar node, and each arm to a sub-band. 
Now we can see that collisions denote instances of two radar nodes utilizing the same spectrum resource, which results in reduced performance. 
As is common, we will assume that collisions are perfectly observable, i.e. each node has some method of detecting collisions. 
Importantly, this does not imply that any node has knowledge of the arms selected by any other node in any time step.

To assign rewards, we will assume that each sub-band has a mean reward in the unit interval which scales with the ambient SINR level. 
In other words the reward observed is dependent on the interference in the selected sub-band. 
Since radar tracking and detection performance scales with SINR, our results show that this reward scenario yields good tracking results for the radar network. 
Specifically rewards will relate to SINR through the relationship in Eq. (\ref{eq:rewards}). 
\begin{equation}
    \label{eq:rewards}
    \mu_{i} = \begin{cases}
    \alpha (SINR+\beta), & c_i(t)=0\\
    0, & c_i(t) = 1
    \end{cases}
\end{equation}
where $\alpha$ and $\beta$ are parameters which bring the various SINR values roughly to the unit interval. 

\subsection{Adversaries} 
We'll consider both adversaries in the traditional sense as well as the case of intelligent emitters. As described above, a traditional adversarial bandit knows the algorithm being implemented by the learner and can specify a reward sequence for each arm \emph{before the game begins}, i.e. Eq. (\ref{eq:re_seq}). 
\begin{equation}
    \label{eq:re_seq}
    \mu_a = \{\mu_a(1), \mu_a(2), \dots, \mu_a(T)\}
\end{equation}
In practice we will restrict the class of environments to those with rewards that change infrequently. 
This mimics a radar environment, which may include interferers which enter and leave the environment unpredictably.

When we consider intelligent interference, we will consider a learner $s$ which in time step $t$ selects action $a^{(s)}(t) = a^{(r)}(t-1)$, which is the action taken in the previous time step by node $r$ in the network. 
We will see that this class of interference causes poor network performance, but the adversarial bandit is still able to outperform the other algorithms we examine. 
This is because with an intelligent emitter, the environment violates the assumption that the reward sequence is decided before the game begins. 
Instead, it depends on the actions chosen by the network.

We simulate the presence of this emitter by changing rewards based on the network's decisions. 
If sub-band $a$ is chosen in time step $t$ and has a mean reward of $\mu_a(t)$, it will have a mean reward of $0$ in time step $t+1$.

\subsection{Algorithms}
\subsubsection{Sense \& Avoid}
We will first build up an algorithm that allows each player in the network to react in a rudimentary manner. Sense \& Avoid (SAA) uses the following algorithm: 
\begin{algorithm}
    \SetAlgoLined
    \KwResult{$a_i(t+1)$} 
    Input $y_i(t), c_i(t)$\\
    \vspace{3mm}
        \eIf{$c_i(t) = 1$}{
        $a_i(t+1) = U(A\backslash a_i(t)O$\\
        }{
        $a_i(t+1) = a_i(t)$
        }
    \caption{Sense \& Avoid}
    \label{algo:SAA}
\end{algorithm}

With perfect collision observation, SAA will quickly converge to a sub-band assignment which does not incur further collisions. 
However, we can not guarantee selection of sub-bands with high rewards. 
A better algorithm would have two attributes: 1) the ability to monitor observed rewards over time and select sub-bands which have consistently high rewards (and correspondingly high target tracking accuracy), and 2) a reward memory length which allows for changing arm rewards. 

\subsubsection{Musical Chairs}
A step in the right direction is the Musical Chairs \cite{pmlr-v48-rosenski16} algorithm, which (in short) allows each player to explore (free from mutual collisions), then come to an agreement on the best $N$ arms in $\mathcal{A}$. 
Lastly, the players will each select one of these $N$ arms to play for the rest of the game. 
\begin{algorithm}
    \SetAlgoLined
    \KwResult{$a_i(t+1)$} 
    Input $y_i(t), c_i(t)$\\
    \vspace{3mm}
        \uIf{phase=Exploration}{
        $a_i(t+1)$ = randsample(M)\;}
        \uElseIf{phase=Exploitation}{
        $a_i(t+1)$ = randsample(bestArms)\;}
        \uElseIf{phase=Fixed}{
        $a_i(t+1) = a_i(t)$\;}
    \caption{Musical Chairs}
    \label{algo:MC}
\end{algorithm}

Musical Chairs (and other algorithms) consists of multiple \emph{phases}, during which the learner behaves differently. 
For example, Musical Chairs has a well-defined exploration phase which lasts for a preset number of steps, during which the learner takes random actions to attempt to learn the mean arm rewards. 
Then, during the exploitation phase, the learner attempts to fixate on an arm with high average rewards.

\subsubsection{Coordinate \& Play}
C\&P \cite{MultiAdversarial} matches our third criteria for a suitable algorithm. 
Implementing MC to assign player rankings, this algorithm assigns the node ranked ``1'' to coordinate the actions of the others. 
Specifically, the coordinating player samples a \emph{meta-arm} $K$ as Eq. (\ref{eq:meta-arm}) 
\begin{equation}
    \label{eq:meta-arm}
    K = \{a_1, \dots, a_N\} | a_i \in \mathcal{A}, \; a_i \neq a_j
\end{equation}
at the beginning of each block. 
Following this definition, we can see that a meta-arm is a subset of the available arms with no duplication. 
This means that within the main part of the C\&P algorithm, the players will experience no collisions. 
With each player using knowledge of its own ranking and the total number of players, the coordinator can sequentially inform each player of the arm to use during that block. 
Importantly, this knowledge is conveyed using intentional collisions.

C\&P divides time by blocks, sub-blocks, and steps. 
Each sub-block consists of $M$ steps, and each block consists of $B > N-1$ sub-blocks. 
At the beginning of each block, player 1 selects a \emph{meta-arm}, which is drawn as above. 
In sub-block $ \leq n < N-1$, the player ranked 1 (also known as the coordinator) uses implicit collisions to give player $n+1$ an arm to use for the remainder of the block. 
This coordination process lasts $N-1$ sub-blocks. 
In the remainder, each player selects the indicated arm. 
Over time, player 1 prefers to pick meta-arms with consistently high rewards. 
Importantly, there is no need for any player to communicate to player 1 the rewards observed, since over time, player 1 will have exploration opportunity. 
This process only incurs collisions (therefore noisy or missing PRIs) on a predictable interval, since the game has repeating cycles.

In sub-block $i$, the coordinator selects arm $a_{i+1}$ until it experiences a collision. 
Correspondingly, in the same sub-block $i$, all players not ranked $i$ will rest, while the player ranked $i$ will select each arm in some order. 
In this manner only one collision will occur per sub-block, and it will be between the coordinator and player $i$ on arm  $a_i$, which player $i$ will then choose for the rest of sub-block $i$ and all sub-blocks $j > N$. 

\begin{algorithm}
    \SetAlgoLined
    \KwResult{$a_i(t+1)$} 
    Input $y_i(t), c_i(t)$\\
    \vspace{3mm}
        \If{t=1}{
        $K \sim \binom{M}{N}$ \# Block begins; sample meta-arm\\}
        \eIf{sub-block $\leq N-1$}{
        $a_i(t+1) = K(\text{sub-block+1})$\\
        }{
        $a_i(t+1) = K(1)$ 
        }
    \caption{Coordinate \& Play - Coordinator}
    \label{algo:CP1}
\end{algorithm}
\begin{algorithm}
    \SetAlgoLined
    \KwResult{$a_i(t+1)$} 
    Input $y_i(t), c_i(t)$\\
    \vspace{3mm}
        \uIf{sub-block+1==ID}{
            \eIf{$c_i(t) \text{ OR } flag=1$}{
            $a_i(t+1) = a_i(t)$ \\
            $A=a_i(t)$ \\
            flag=1
            }{
            $a_i(t+1) = t\mod M$\\
            }
        }
        \uElseIf{sub-block $<$ N-1}{
            $a_i(t+1) = 0$ \;
        }
        \Else{
            $a_i(t+1) = A$ \;
        }
    \caption{Coordinate \& Play - Follower}
    \label{algo:CP2}
\end{algorithm}

We provide a sketch of the coordinator algorithm in Algorithm \ref{algo:CP1}, and of the follower algorithm in Algorithm \ref{algo:CP2}. For full details see \cite{MultiAdversarial}.

With this algorithm in mind, we can tune our PRI length to suite the time increments described above.  
Since the coordination period occurs at the beginning of each block, we'd like there to be an integer number of CPI's per block.
% If it were otherwise, then every single CPI will include PRIs with intentionally poor performance. 
% If we can limit the frequency of this behavior to once every several CPIs, we can asymptotically guarantee optimal CPIs since the algorithm will converge towards the optimal solution. 

We select the block size as \cite{MultiAdversarial}
\begin{equation}
\label{eq:block_size}
    \tau = \left(\frac{K^2NT}{\log(N)}\right)^{\frac13}
\end{equation}
where $K$ is the number of nodes, $N$ is the number of sub-bands, and $T$ is some finite time horizon. 
Note that in addition to selecting block size, we specify the size of each sub block as containing $N$ steps. This is the minimum number of steps necessary to guarantee implicit communication from the coordinator to a given follower.

Using a network of three nodes and five actions, with a time horizon of $10^4$, we arrive at a block size of $48$ which we round to $50$. 
%%%%%%%%%%%%%%%%%%%%%%%%%%%%%%%%%%%%%%%%%%%%%%%%%%%%%%%%%%%%%%%%%%%%%%%%%%%%%%%%%%%%%%%%%

%%%%%%%%%%%%%%%%%%%%%%%%%%%%%%%%%%%%%%%%%%%%%%%%%%%%%%%%%%%%%%%%%%%%%%%%%%%%%%%%%%%%%%%%%
\section{Results}
Previous work \cite{Howard_multiplayer} has considered the interactions of radar networks in environments which do not vary over time. 
In other words, the reward distribution for a given action is independent of the time step. 
Adversarial environments are those which can vary over time according to a sequence which is selected \emph{before the game begins}. 
Environments with non-reactive emitters fall under this category. 
We will consider such an environment, where one or more unknown emitters access the environment according to some pre-determined, but \emph{unknown to the radar network} sequence.

In the following simulations, rewards are drawn from Gaussian distributions with means $\mu_1 = 0.95, \mu_2 = 1, \mu_3 = 0.9, \mu_4 = 0.3, \mu_5 = 0.3$. 
After time step 200, the means shift to $\mu_1 = 0.3, \mu_2 = 0.3, \mu_3 = 0.95, \mu_4 = 1, \mu_5 = 0.9$. 
We simulate 500 CPIs, each of which contains 50 PRIs. 
For Coordinate \& Play, we use a block size of 10 sub-blocks, and a sub-block size of 5 PRIs as described above. 
Results shown below are averaged over 50 independent trials unless otherwise stated. 

The radar scenario shown demonstrates a target moving from the origin to $[350, 100]$ as a network consisting of three cognitive radar nodes attempts to make tracking estimates. 
The radar nodes are located at $[0, 500]$, $[250, -100]$, and $[500, 500]$ and do not move. 

\begin{figure}[htbp]
    \centering
    \includegraphics[scale=0.6]{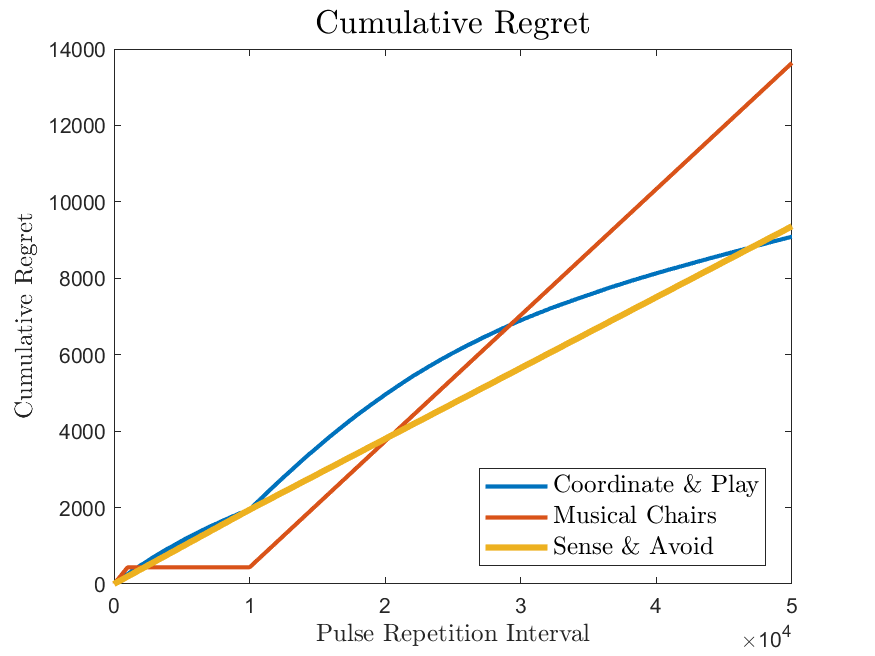}
    \caption{Cumulative regret for three cognitive radar networks. }
    \label{fig:cum_regret}
\end{figure}

In Fig. \ref{fig:cum_regret} we show the regret of two algorithms: Musical Chairs and Coordinate \& Play. 
We can see that during the first 200 CPIs, when the rewards remain constant, Musical Chairs outperforms Coordinate \& Play. 
However, once the environment shifts rewards, we can see that Coordinate \& Play is able to find the new optimum over time. 
Musical Chairs has no ability to continue monitoring the environment once the players fixate on a solution. 
Coordinate \& Play however monitors the rewards observed over time, and can take more exploratory actions if the rewards start to change. 

\begin{figure}[htbp]
    \centering
    \includegraphics[scale=0.6]{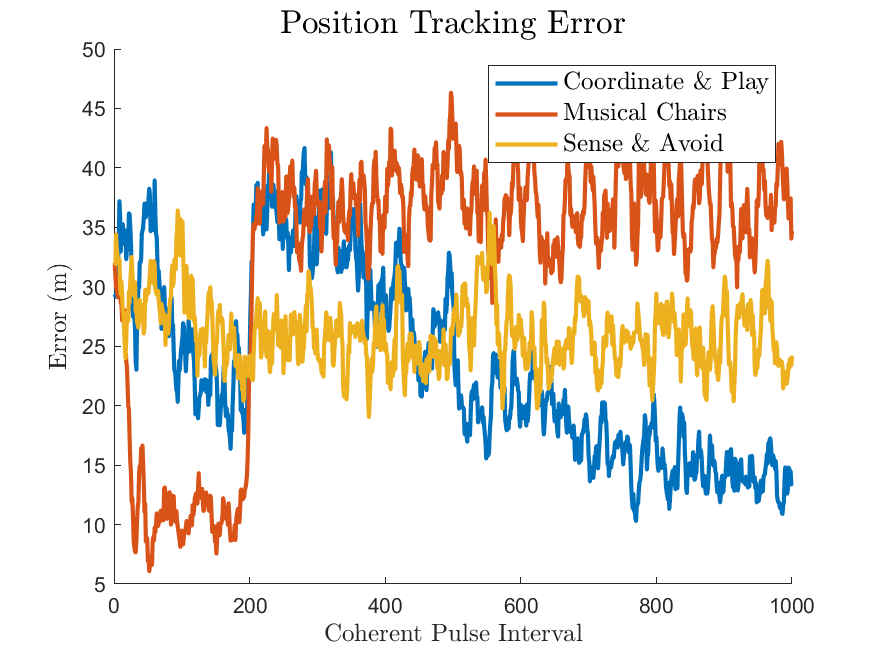}
    \caption{A comparison of the position estimation error for three networks using different algorithms: Coordinate \& Play, Musical Chairs, and Sense \& Avoid. After the 200$^{th}$ Coherent Pulse Interval, the environment shifts and provides different rewards. The Coordinate \& Play network is able to converge to the new optimum solution, thus maintaining tracking accuracy, while the other two networks do not. }
    \label{fig:error}
\end{figure}

Fig. \ref{fig:error} shows that radar network tracking performance corresponds to regret. 
Note that Fig.\ref{fig:cum_regret} is in units of Pulse Repetition Interval, since we can calculate rewards for each pulse, while Fig. \ref{fig:error} is in units of Coherent Pulse Intervals, since we only make tracking estimates after each CPI. 
We can see that the Musical Chairs network performs similarly to the C\&P network for the first 200 CPIs, but then fails to recover once the environment shifts. 
On the other hand, the C\&P network performs worse after the environment shifts, but is able to adapt and maintain tracking performance. 
We can also see that Sense \& Avoid quickly converges to a solution with no mutual collisions, but with a high regret bound. 
This occurs because the only intelligence Sense \& Avoid has is to avoid other nodes. 
Once no more collisions occur, no nodes will change their sub-band. 
We can see that this approach performs worse than Musical Chairs, since before the environment shifts, Musical Chairs is at least able to converge to a good solution. 
\begin{figure}[htbp]
    \centering
    \includegraphics[scale=0.6]{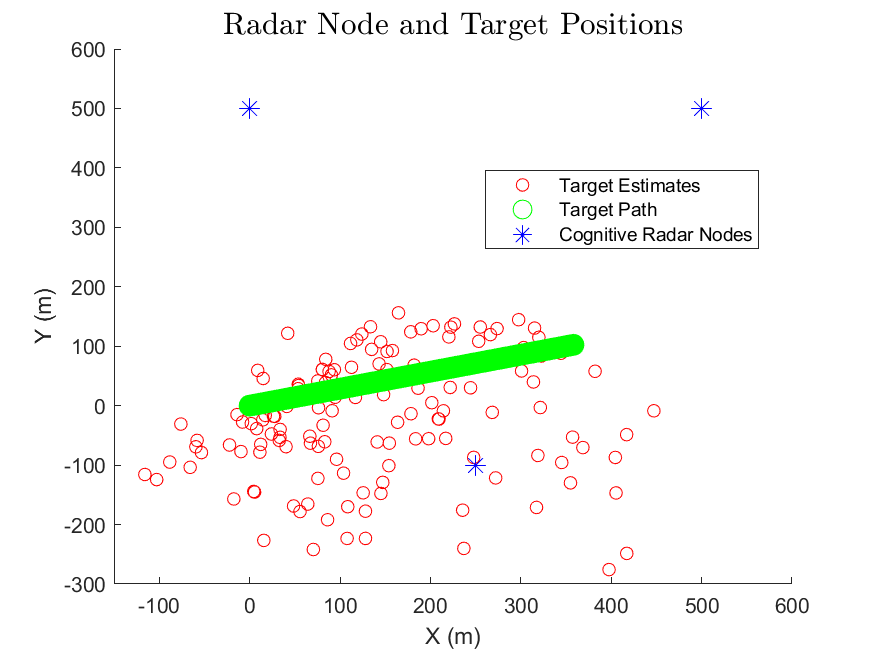}
    \caption{Estimated target positions for a single simulation. }
    \label{fig:positions}
\end{figure}

Fig. \ref{fig:positions} shows the results of a single simulation. We see that many target position estimates have relatively high error, but these constitute a small fraction of the overall estimates.

\begin{figure}[htbp]
    \centering
    \includegraphics[scale=0.6]{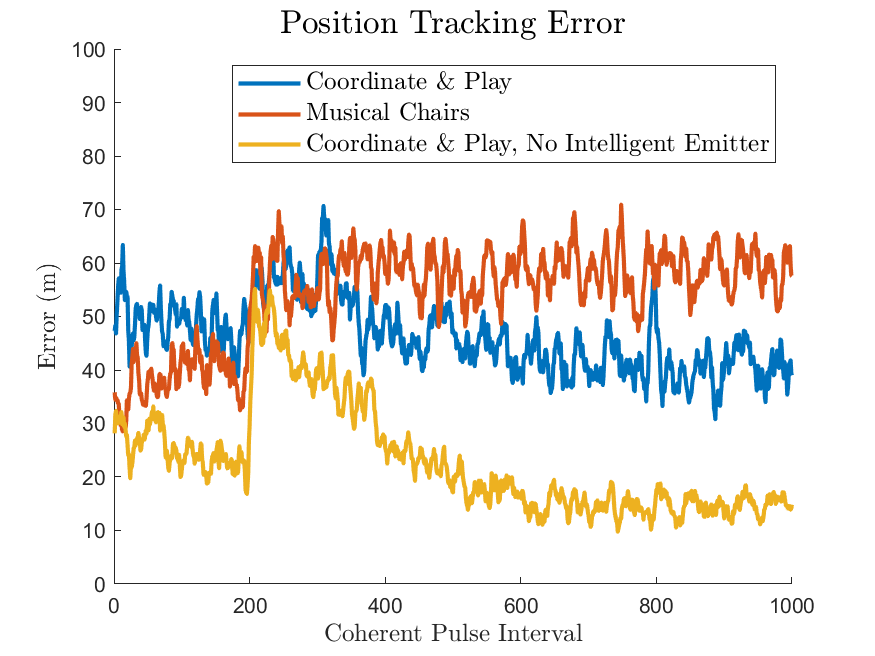}
    \caption{Positioning error for three networks: C\&P with an intelligent emitter, Musical Chairs with an intelligent emitter, and C\&P without any other emitters. We can see that the intelligent emitter does causes the C\&P network to perform worse initially than the Musical Chairs network, but after the environment shifts, the C\&P network begins to outperform the Musical Chairs network. Naturally, both of these underperform when compared to the C\&P network without other emitters. }
    \label{fig:FollowError}
\end{figure}
In Fig. \ref{fig:FollowError}, we see a comparison of three different networks. Two are impaired by an intelligent emitter, while the third is not. We can see that the Coordinate \& Play algorithm performs worse under this environment than under the adversarial environment as defined above. 
This is because the intelligent emitter is able to observe the previous center frequency of a node, then select that sub-band in the subsequent PRI. 
This breaks the assumption that an adversarial environment is unable to modify its loss sequence after the game begins; this emitter causes the loss to react each PRI. 

\begin{figure}[htbp]
    \centering
    \includegraphics[scale=0.6]{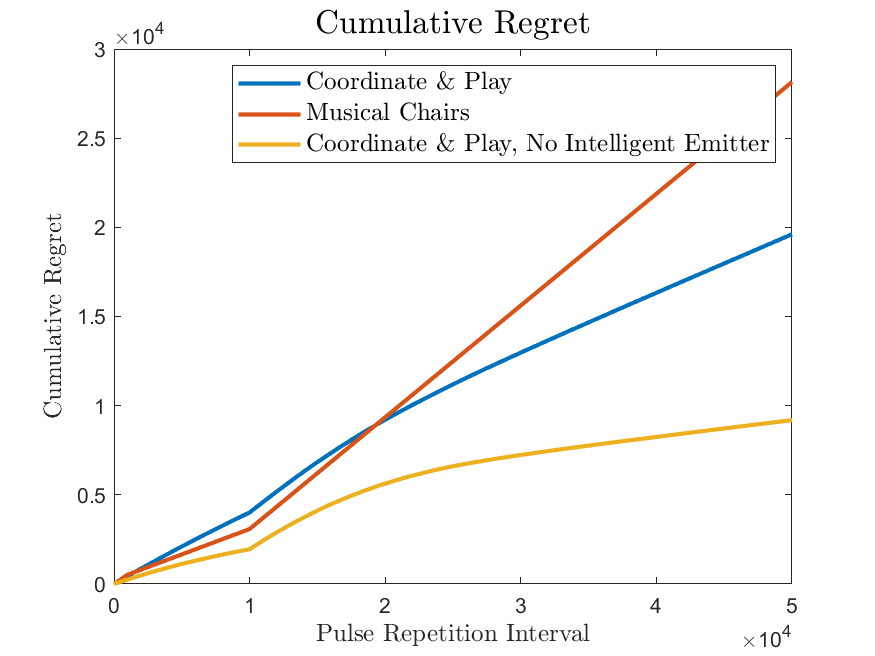}
    \caption{Regret for the same three networks as in Fig. \ref{fig:FollowError}. }
    \label{fig:FollowRegret}
\end{figure}
%%
%%%%%%%%%%%%%%%%%%%%%%%%%%%%%%%%%%%%%%%%%%%%%%%%%%%%%%%%%%%%%%%%%%%%%%%%%%%%%%%%%%%%%%%%%

%%%%%%%%%%%%%%%%%%%%%%%%%%%%%%%%%%%%%%%%%%%%%%%%%%%%%%%%%%%%%%%%%%%%%%%%%%%%%%%%%%%%%%%%%
\section{Conclusions}
We have shown the design of an adversarial radar environment where different sub-bands have SINR values that change without notice. In this environment, a radar network with static frequency allocations, or even one implementing an algorithm capable of selecting the initial optimal sub-bands, will tend to fail. However, those using adversarial multi-player bandit algorithms are more able to deal with this varying environment.

When we consider environments with some memory length and ability to react to the actions of the radar network, we can see that the adversarial bandit formulation breaks down but is still able to outperform less dynamic algorithms such as Musical Chairs and Sense \& Avoid. 
This highlights the fact that understanding the class of environment is paramount to consistent radar performance. 
If the algorithm is tailored to a static environment and the SINR values suddenly change, that network may no longer obtain high accuracy tracking estimates.

Through the use of adversarial multi-player multi-armed bandit algorithms, a cognitive radar network can continue to track targets while maintaining accuracy in a changing environment. 
%%%%%%%%%%%%%%%%%%%%%%%%%%%%%%%%%%%%%%%%%%%%%%%%%%%%%%%%%%%%%%%%%%%%%%%%%%%%%%%%%%%%%%%%%

% \clearpage
\newpage
\bibliographystyle{IEEEtran}
\bibliography{bibli}
\end{document}